\documentclass[aps,nofootinbib,11pt]{revtex4-1}
\usepackage[letterpaper,hmargin=1in,vmargin=1in]{geometry}

\usepackage{graphicx,epstopdf,amsmath,amsfonts,amssymb,amsxtra,color}

\usepackage{multirow,array,booktabs}
\usepackage{cancel,bbold,extarrows}
\usepackage{hyperref,cleveref}

\usepackage{aas_macros} 

\begin{document}
 
\title{Constraints on force-free magnetospheres for Kerr(-AdS) black holes with non-null currents}

\author{Xun Wang}
\affiliation{Department of Physics and Astronomy, University of Victoria, 
     Victoria, BC, V8P 5C2 Canada}

\date{May 2015}

\begin{abstract}
\noindent

Force-free magnetospheres are of particular interest due to their role in energy extraction from Kerr black holes via the Blandford-Znajek process. Recently, a class of exact analytic solutions has been found with null currents \cite{MenonDermer,Jacobson}. In this paper, we elaborate some constraints on various force-free magnetosphere solutions with non-null currents, utilizing the Newman-Penrose electromagnetic scalars to categorize a range of different cases. We perform a thorough search for stationary and axisymmetric (SAS) solutions, and find that putative SAS solutions within the categories considered generically exhibit singularities on the horizon. We also present some non-SAS solutions found via spacetime-dependent electric-magnetic duality rotations. Additional special solutions in flat, pure AdS and near-horizon-extreme-Kerr (NHEK) spacetimes are also presented.

\end{abstract}
\maketitle

\section{Introduction}

Force-free magnetospheres around rotating Kerr black holes allow for the extraction of rotational energy in the form of a Poynting flux, known as the Blandford-Znajek (BZ) process \cite{BZ}. Such configurations are of particular physical interest as they may provide the primary power source for active galaxies. Blandford and Znajek originally obtained solutions in the slow rotation limit, but physical applications motivate the search for general solutions, particularly in the background of near-extremal black holes. It has proven to be challenging to find exact analytic solutions, but nevertheless various special solutions have recently emerged. Examples include those with the current aligned along a principle null direction of the Kerr geometry \cite{MenonDermer,Jacobson}, solutions imposing translational (rather than axi-) symmetry \cite{Jac_jets}, and those making use of the extra symmetries of the near-horizon extreme Kerr (NHEK) background \cite{Strominger:NHEK_force-free,Lehner:NHEK_force-free} (also \cite{Li:nearNHEK}). It is also worth mentioning some new approximate solutions; see \cite{4thBZ,Gralla:jets} and references therein.

Formally, the basic \emph{force-free equations} (conservation equations for the electromagnetic energy-momentum tensor) read
\begin{equation}\label{ff}
T_{\mu;\nu}^\nu=F_{\nu\mu}J^\nu=0,
\end{equation}
which implies the \emph{degeneracy condition} on the electromagnetic field
\begin{equation}\label{degen}
I_2\equiv{^\star}F_{\mu\nu}F^{\mu\nu}=0,
\end{equation}
due to the fact that $I_2\propto\det F_{\mu\nu}=0$ for non-zero currents $J^\mu$. The sign of the other invariant $I_1\equiv F_{\mu\nu}F^{\mu\nu}$ indicates whether the configuration is magnetically dominant ($I_1>0$), electrically dominant ($I_1<0$) or null ($I_1=0$).

Here the current is simply defined via $J^\mu\equiv F^{\mu\nu}_{;\nu}$, rather than being a prescribed physical source. Nevertheless, by making additional assumptions about certain properties of $J^\mu$ one can hope to simplify the equations, along with stronger constraints than \eqref{degen} on the field $F_{\mu\nu}$. E.g., the null current assumption in \cite{MenonDermer,Jacobson} leads to the extra constraint $\phi_1=\phi_2=0$ or $\phi_1=\phi_0=0$, with $F_{\mu\nu}$ given equivalently by three complex Newman-Penrose (NP) variables $\{\phi_0,\phi_1,\phi_2\}$ (reviewed below). In this paper, we classify NP variables directly, making use of a Kinnersley-like null tetrad, and check properties of the current for each solution. Our main results are for the case with only $\phi_1=0$, including SAS (stationary and axisymmetric) and non-SAS solutions.

For SAS solutions, we give a more thorough analysis. In fact, one can show that among the cases where at least one NP variable vanishes, the null current case ($\phi_1=\phi_2=0$ or $\phi_1=\phi_0=0$) and the case $\phi_1=0$ are of most interest; other cases either reduce to them or have no solutions (we show this in the appendix). We then turn to more general cases where all NP variables are non-zero: first the cases with $\Im(\phi_1^2)=0$ (including $\Re\phi_1=0$ and $\Im\phi_1=0$) and then the remaining ones with $\Im(\phi_1^2)\neq0$ (classified according to $\Re(\phi_1^2)$). Our primary observation is that putative solutions in these restricted cases have a generic on-horizon singularity unless the current is null.

The non-SAS solutions are found via spacetime-dependent electric-magnetic \emph{duality rotations} (defined later on) which leave the force-free equations invariant but map the field and current to different configurations. Using such rotations as a solution generating technique, we find that some of the null and non-null current solutions can be ``duality-rotated'' to a vacuum solution, providing an interesting class of dual configurations. We observe that requiring stationarity and axisymmetry (SAS) on both sides seems too restrictive (the duality rotation mixes $F_{\mu\nu}$ and its dual), and time/azimuthally-dependent solutions usually result.

The paper is organized as follows. In \cref{sec_ff}, we reformulate the force-free equations using the NP formalism and introduce some convenient notation. After briefly reviewing the previously obtained null current solutions in \cref{sec_Jac}, we present special solutions with non-null currents in \cref{phi1=0}, concentrating on the SAS case. In \cref{sec_rot}, we use the spacetime-dependent duality rotation to find pairs of solutions with one force-free and the other a vacuum configuration, for both null and non-null currents. In addition, some special SAS solutions are found for flat, AdS and NHEK spacetimes in \cref{sec_SAS}. A discussion of the results is given in \cref{disc}. Throughout, we consider generalizations of Kerr to Kerr-AdS geometries, having in mind potential applications of the AdS/CFT correspondence to provide an alternate dual description of these solutions, as studied in \cite{paper1}.

\section{Force-free equations in Newman-Penrose (NP) formalism}\label{sec_ff}

\subsection{formulation of the force-free equations}

We write the general metric in $3+1$ formalism \cite{3+1Formalism} as
\begin{equation}\label{3+1metric}
{\operatorname{d}\!s}^2=-\alpha^2{\operatorname{d}\!t}^2+h_{ij}\bigl(\operatorname{d}\!x^i+\beta^i\operatorname{d}\!t\bigr)\bigl(\operatorname{d}\!x^j+\beta^j\operatorname{d}\!t\bigr),\qquad (i,j=\text{spatial directions}).
\end{equation}
The Kerr-AdS metric in Boyer-Lindquist (BL) coordinates is given by
\begin{align}
h_{rr}&=\frac{\Sigma}{\Delta_r},\qquad h_{\theta\theta}=\frac{\Sigma}{\Delta_\theta},\qquad h_{\varphi\varphi}=\frac{\Delta_\theta(r^{2}+a^{2})^{2}-\Delta_ra^{2}\sin^{2}\theta}{\Xi^{2}\Sigma}\sin^{2}\theta\\
\alpha^{2}&=\frac{\Sigma\Delta_r\Delta_\theta}{\Delta_\theta(r^{2}+a^{2})^{2}-\Delta_ra^{2}\sin^{2}\theta}, \qquad
\beta^i=\Bigl[0,0,-a\Xi\frac{\Delta_\theta(r^{2}+a^{2})-\Delta_r}
{\Delta_\theta(r^{2}+a^{2})^{2}-\Delta_ra^{2}\sin^{2}\theta}\Bigr],
\end{align}
where
\begin{align}
\Sigma=r^{2}+a^{2}\cos^{2}\theta, \qquad &\Xi =1-\frac{a^2}{l^2}\\
\Delta_r=(r^{2}+a^{2})\bigl(1+\frac{r^2}{l^2}\bigr)-2mr,& \qquad
\Delta_\theta=1-\frac{a^2}{l^2}\cos^{2}\theta,
\end{align}
and $\sqrt{-g}=\Sigma\sin\theta/\Xi$. The Kerr limit is $l\rightarrow\infty$; especially, $\Xi,\Delta_\theta\rightarrow1$ and $\Delta_r\rightarrow\Delta=r^2+a^2-2mr$.

In the Newman-Penrose (NP) formalism \cite{NP_paper,ChandrasekharBH,ExactSolns}, the electromagnetic field $F_{\mu\nu}$ is specified by the following three complex NP variables (bar denoting complex conjugate)
\begin{align}
\phi_0&\equiv F_{\mu\nu}l^\mu m^\nu\\
\phi_1&\equiv \frac{1}{2}F_{\mu\nu}(l^\mu n^\nu+\bar m^\mu m^\nu)\\
\phi_2&\equiv F_{\mu\nu}\bar m^\mu n^\nu.
\end{align}
For Kerr-AdS, we use the following Kinnersley-like null tetrad \cite{Kerr-AdS_tetrad} \footnote{This tetrad corresponds to the metric signature $[+,-,-,-]$. Also, the $t$-components differ from those in \cite{Kerr-AdS_tetrad} by a factor of $\Xi$.}
\begin{align}
l^\mu&=\Bigl[\frac{r^2+a^2}{\Delta_r},1,0,\frac{a\Xi}{\Delta_r}\Bigr]\\
n^\mu&=\frac{\Delta_r}{2\Sigma}\Bigl[\frac{r^2+a^2}{\Delta_r},-1,0,\frac{a\Xi}{\Delta_r}\Bigr]\\
m^\mu&=-\bar\rho\sqrt{\frac{\Delta_\theta}{2}}\Bigl[\frac{ia\sin\theta}{\Delta_\theta},0,1,\frac{i\Xi}{\Delta_\theta\sin\theta}\Bigr],
\end{align}
listing components in the order $[t,r,\theta,\varphi]$, where $\rho\equiv-(r-ia\cos\theta)^{-1}$. Note that $\rho\bar\rho=1/\Sigma$. The tetrad vectors are normalized as $l^\mu n_\mu=-m^\mu\bar m_\mu=1$ so that the metric is $g_{\mu\nu}=2l_{(\mu}n_{\nu)}-2m_{(\mu}\bar m_{\nu)}$.
We use indices in parentheses $\{(1),(2),(3),(4)\}$, interchangeably with $\{l,n,m,\bar m\}$ (as indices), to indicate tensor components from contractions with $\{l^\mu,n^\mu,m^\mu,\bar m^\mu\}$ respectively. The currents are (dropping a factor of 2)
\begin{align}
J_l&=\rho^2\nabla_l\phi_1'-\frac{1}{\rho\Delta_r\sqrt{\Delta_\theta}\sin\theta}\nabla_{\bar m}\phi_0'\label{J_l}\\
J_n&=-\rho^2\nabla_n\phi_1'+\frac{\rho}{\sqrt{\Delta_\theta}\sin\theta}\nabla_m\phi_2'\label{J_n}\\
J_m&=\rho^2\nabla_m\phi_1'-\frac{1}{\rho\Delta_r\sqrt{\Delta_\theta}\sin\theta}\nabla_n\phi_0'\label{J_m}\\
J_{\bar m}&=-\rho^2\nabla_{\bar m}\phi_1'+\frac{\rho}{\sqrt{\Delta_\theta}\sin\theta}\nabla_l\phi_2',\label{J_mbar}
\end{align}
where we have introduced the rescaled NP variables
\begin{equation}\label{phi'}
\phi_0'\equiv\rho\Delta_r\sqrt{\Delta_\theta}\sin\theta\phi_0,\qquad\phi_1'\equiv\rho^{-2}\phi_1,\qquad\phi_2'\equiv\rho^{-1}\sqrt{\Delta_\theta}\sin\theta\phi_2.
\end{equation}
The above `definitions' of $J_{(a)}$ have incorporated the homogeneous Maxwell equations $\operatorname{d}\!F=0$ which read
\begin{equation}\label{dF=0_NP_J}
J_l=\bar J_l,\qquad J_n=\bar J_n,\qquad J_m=\bar J_{\bar m},\qquad J_{\bar m}=\bar J_m.
\end{equation}
Later when we present special solutions for $\phi_{0,1,2}$, we will need to verify \eqref{dF=0_NP_J} separately. The force-free equations $F_{(a)(b)}J^{(b)}=0$ are \cite{Jacobson}
\begin{equation}\label{FJ_eqns}
\Re(\phi_1J_n-\phi_2J_m)=0,\qquad \Re(\phi_1J_l-\bar\phi_0J_m)=0,\qquad 2i\Im\phi_1J_{\bar{m}}+\bar\phi_0J_n-\phi_2J_l=0.
\end{equation}
The degeneracy condition is $\Im(\phi_0\phi_2-\phi_1^2)=0$ while for the invariants we have $I\equiv I_1+iI_2=8(\phi_0\phi_2-\phi_1^2)$. ``$\Re,\Im$'' denote real and imaginary parts.

When appropriate, we also express some results in terms of the \emph{modified} NP variables (with the same $\phi_1$):
\begin{equation}\label{Phi1Phi2}
\Phi_1\equiv\Delta_r\rho\phi_0+\frac{2\phi_2}{\rho},\qquad \Phi_2\equiv\Delta_r\rho\phi_0-\frac{2\phi_2}{\rho},\qquad
\Phi_{1,2}=\frac{\phi_0'\pm2\phi_2'}{\sqrt{\Delta_\theta}\sin\theta},
\end{equation}
the advantage of which being that we can transform to the real electromagnetic field components more easily:
\begin{align}
F_{r\varphi}&=-2\frac{a\sin^2\theta}{\Xi}\Re\phi_1-\frac{(r^2+a^2)\sin\theta\sqrt{\Delta_\theta}}{\sqrt2\Xi\Delta_r}\Im\Phi_2\label{F_rph_NP}\\
F_{rt}&=2\Re\phi_1+\frac{a\sin\theta\sqrt{\Delta_\theta}}{\sqrt2\Delta_r}\Im\Phi_2\label{F_rt_NP}\\
F_{\theta\varphi}&=-\frac{a\sin^2\theta}{\sqrt2\Xi\sqrt{\Delta_\theta}}\Re\Phi_2+2\frac{(r^2+a^2)\sin\theta}{\Xi}\Im\phi_1\label{F_thph_NP}\\
F_{\theta t}&=-2a\sin\theta\Im\phi_1+\frac{\Re\Phi_2}{\sqrt2\sqrt{\Delta_\theta}}\label{F_tht_NP}\\
B_T+iF_{\varphi t}&=\frac{\sin\theta\sqrt{\Delta_\theta}}{\sqrt2\Xi}\Phi_1\label{BT_Eph_NP},
\end{align}
where
\begin{equation}
B_T\equiv-\alpha^2h_{\varphi\varphi}B^\varphi=-\frac{\Delta_r\Delta_\theta}{\Xi^2}\sin^2\theta B^\varphi,\qquad \text{with }B^\varphi=F_{r\theta}/\sqrt{-g}.
\end{equation}
For stationary and axisymmetric configurations where $F_{\varphi t}$ vanishes, the degeneracy condition $I_2=0$ implies the existence of the ratio $\omega=-F_{tr}/F_{\varphi r}=-F_{t\theta}/F_{\varphi\theta}$ which represents the ``angular velocity'' of the magnetic field lines.  For completeness, the force-free equations in terms of $\{\Phi_{1,2},\phi_1\}$ are
\begin{align}
(\Im\Phi_1+\Im\Phi_2)J_A^P-2\bar\rho\Delta_r\Im\phi_1J_B^P&=0\label{FJ_ab_1}\\
(\Re\Phi_2+i\Im\Phi_1)J_A^T+\bar\Phi_1J_A^P+2i\bar\rho\Delta_r\Im\phi_1J_B^T&=0,\label{FJ_ab_2}
\end{align}
where \footnote{The superscript ``$T/P$'' indicates that, e.g., $J_A^{T/P}$ only involves the contractions of $J_\mu$ with the toroidal/poloidal components of $n^\mu$. Note that in the Kerr limit, $\Phi_1$ \& $\Phi_2$ are proportional to the electric and magnetic field components in the orthonormal frame associated with the Carter tetrad \cite{Carter_tetrad}, e.g., $\Phi_1\propto(E_3+iB_3)_\text{Carter}$ etc., and $J^{T/P}_{A/B}$ proportional to components of the current in the same orthonormal frame.}
\begin{equation}
2J_A^{T/P}\equiv J_n\pm\frac{\Delta_r}{2\Sigma}J_l,\qquad 2J_B^{T/P}\equiv J_{\bar m}\mp\frac{\rho}{\bar\rho}J_m.
\end{equation}

\subsection{Regularity conditions on the horizon and at the poles}\label{reg_conditions}

Let's pause to clarify regularity conditions in terms of NP variables. On the horizon $\Delta_r=0$, the field components should be regular in the following (Kerr-AdS analog of) ingoing Kerr (IK) coordinates defined via
\begin{equation}\label{ingoingKerr}
\operatorname{d}\!\psi=\operatorname{d}\!\varphi+\frac{a\Xi}{\Delta_r}\operatorname{d}\!r,\qquad
\operatorname{d}\!v=\operatorname{d}\!t+\frac{r^2+a^2}{\Delta_r}\operatorname{d}\!r,
\end{equation}
in which the metric takes the form
\begin{equation}
\operatorname{d}\!s^2 =-\frac{\Delta_r}{\Sigma}\Bigl[\operatorname{d}\!v-\frac{a}{\Xi} \sin^2\theta\operatorname{d}\!\psi-\frac{\Sigma}{\Delta_r}\operatorname{d}\!r\Bigr]^2+\frac{\Sigma}{\Delta_r}\operatorname{d}\!r^2+\frac{\Sigma}{\Delta_\theta}\operatorname{d}\!\theta^2+\frac{\Delta_\theta\sin^2\theta}{\Sigma}\Bigl[\frac{r^2+a^2}{\Xi}\operatorname{d}\!\psi-a\operatorname{d}\!v\Bigr]^2,
\end{equation}
the only difference being the new $\operatorname{d}\!r$ term in the first bracket (apart from replacing $(\varphi,t)\rightarrow(\psi,v)$). In particular,
\begin{align}
F^\mathrm{IK}_{r\theta}&=F_{r\theta}+\partial_rt(v,r)F_{t\theta}+\partial_r\varphi(\psi,r)F_{\varphi\theta}\\
&=-\frac{\Sigma}{\sqrt2\Delta_r\sqrt{\Delta_\theta}}(\Re\Phi_1-\Re\Phi_2)=-\frac{2\sqrt2\Sigma}{\Delta_r\Delta_\theta\sin\theta}\Re\phi_2'\label{F_rth_IK}\\
F^\mathrm{IK}_{r\varphi}&=F_{r\varphi}+\partial_rt(v,r)F_{t\varphi}\\
&=-2\frac{a\sin^2\theta}{\Xi}\Re\phi_1+\frac{(r^2+a^2)\sin\theta\sqrt{\Delta_\theta}}{\sqrt2\Xi\Delta_r}(\Im\Phi_1-\Im\Phi_2)=-2\frac{a\sin^2\theta}{\Xi}\Re\phi_1+2\sqrt2\frac{r^2+a^2}{\Xi\Delta_r}\Im\phi_2'\label{F_rph_IK}\\
F^\mathrm{IK}_{rt}&=F_{rt}+\partial_r\varphi(\psi,r)F_{\varphi t}\\
&=2\Re\phi_1-\frac{a\sin\theta\sqrt{\Delta_\theta}}{\sqrt2\Delta_r}(\Im\Phi_1-\Im\Phi_2)=2\Re\phi_1-2\sqrt2\frac{a}{\Delta_r}\Im\phi_2'.\label{F_rt_IK}
\end{align}
Other components are not affected by the coordinate transformations. (Note that in BZ's original monopole ansatz, $F_{r\theta}^\mathrm{IK}=0$ provides exactly the horizon regularity condition for $F_{r\theta}\propto B_T$.)

For regularities at the poles, we use the Cartesian Kerr-Schild coordinates $[\tau,x,y,z]$ \cite{HawkingEllis,MenonDermer_singular}, defined by (for simplicity consider the Kerr limit)
\begin{equation}
\tau=v-r,\qquad x=(r\cos\psi-a\sin\psi)\sin\theta,\qquad y=(r\sin\psi+a\cos\psi)\sin\theta,\qquad z=r\cos\theta,
\end{equation}
in which the Kerr metric becomes \cite{HawkingEllis}
\begin{equation}
\operatorname{d}\!s=\operatorname{d}\!x^2+\operatorname{d}\!y^2+\operatorname{d}\!z^2-\operatorname{d}\!\tau^2+\frac{2mr^3}{r^4+a^2z^2}\Bigl(\frac{r(x\operatorname{d}\!x+y\operatorname{d}\!y)-a(x\operatorname{d}\!y-y\operatorname{d}\!x)}{r^2+a^2}+\frac{z\operatorname{d}\!z}{r}+\operatorname{d}\!\tau^2\Bigr)^2
\end{equation}
and $\sqrt{-g}=1$. Components of physical quantities include
\begin{align}
J^\tau&=J^v-J^r,\qquad J^z=\cos\theta J^r-r\sin\theta J^\theta\\
J^x&=\cos\psi\sin\theta J^r+(r\cos\psi-a\sin\psi)\cos\theta J^\theta-(r\sin\psi+a\cos\psi)\sin\theta J^\psi\\
J^y&=\sin\psi\sin\theta J^r+(r\sin\psi+a\cos\psi)\cos\theta J^\theta+(r\cos\psi-a\sin\psi)\sin\theta J^\psi
\end{align}
and similarly for the energy fluxes, with $T_\tau^a$ replacing $J^a$ on the l.h.s and $T_v^\mu$ replacing $J^\mu$ on the r.h.s of the above expressions. It is easy to see that finite $J^z$ can only be realized with $J^r\sim J^\theta\sim\mathcal{O}(1)$ (or more regular). For $F_{\mu\nu}$ the criteria are \cite{MenonDermer_singular}
\begin{equation}\label{poles_F}
F_{\mu\varphi}\sim\mathcal{O}(\sin\theta),\qquad\text{other }F_{\mu\nu}\sim\mathcal{O}(1),
\end{equation}
which translate to (according to \eqref{F_rph_NP}--\eqref{BT_Eph_NP})
\begin{equation}\label{poles_NP}
\Phi_{1,2}\sim\phi_1\sim\mathcal{O}(1),\qquad B_T\sim\mathcal{O}(\sin\theta).
\end{equation}
(For reference, BZ's monopole solution scales as $J^\mu\sim\mathcal{O}(1),F_{\theta\varphi}\sim\mathcal{O}(\sin\theta),B_T\sim\mathcal{O}(\sin^2\theta)$.)

\section{Review of null current solutions}\label{sec_Jac}

We first generalize Brennan et al's null current solutions \cite{Jacobson} in Kerr to Kerr-AdS in a straightforward way, following basically the same arguments. Consider a null current $J^\mu$ along the principle null direction $n^\mu$ so that only the real component $J_l=n_\mu l^\mu\equiv\mathcal{J}$ is non-zero. In the ingoing Kerr coordinates $\nabla_n$ becomes simply $\partial_r$. The whole force-free problem can be shown to reduce to the Maxwell equations
\begin{align}
\sqrt{2}\mathcal{J}&=\sqrt{2}\Re J_l=\frac{1}{\Delta_r\sqrt\Delta_\theta\sin\theta}\bigl(\Xi\partial_\psi\Im\Phi_1+a\sin^2\theta\partial_v\Im\Phi_1\bigr)+\frac{\partial_\theta(\Re\Phi_1\sqrt\Delta_\theta\sin\theta)}{\Delta_r\sin\theta}\label{Jl_eqn_Re}\\
0&=\sqrt{2}\Im J_l=-\frac{1}{\Delta_r\sqrt\Delta_\theta\sin\theta}\bigl(\Xi\partial_\psi\Re\Phi_1+a\sin^2\theta\partial_v\Re\Phi_1\bigr)+\frac{\partial_\theta(\Im\Phi_1\sqrt\Delta_\theta\sin\theta)}{\Delta_r\sin\theta}\label{Jl_eqn_Im},
\end{align}
with $\phi_1=\phi_2=0,\;\Phi_1=\Phi_1(\theta,\psi,v)=\Phi_2$. The Kerr-AdS analog of an SAS solution first found by Menon and Dermer \cite{MenonDermer} and re-derived in \cite{Jacobson} is given as
\begin{equation}\label{S(theta)}
\Phi_1=\Re\Phi_1=\frac{\sqrt{\Delta_\theta}}{\Xi}S'(\theta),\qquad[\sin\theta\Delta_\theta S'(\theta)]'=\sqrt{2}\Xi\Delta_r\sin\theta\mathcal{J}(r,\theta).
\end{equation}
The non-vanishing real field quantities are
\begin{equation}\label{real_null}
B_T=\frac{\sin\theta\sqrt{\Delta_\theta}}{\sqrt2\Xi}\Phi_1,\qquad\omega=\frac{\Xi}{a\sin^2\theta}.
\end{equation}
Later in \cref{rot_null} we will duality-rotate this solution to a vacuum one.

\section{Solutions with non-null currents}\label{phi1=0}

\subsection{Field configurations}\label{classes}

The null current solutions have the common constraints $\phi_1=\phi_2=0$ (or $\phi_1=\phi_0=0$ if we assume $J_{(a)}=J_n$ rather than $J_{(a)}=J_l$), and the electromagnetic field is necessarily ``null'' (by which one means $I_1=I_2=0$). In this section we try to construct some solutions with weaker constraints on the field (so that it is not null) and then infer the causal nature of the currents \emph{a posteriori}. We organize different situations below based on values of NP variables. The search is not exhaustive and is restricted to SAS solutions. These solutions have non-null currents and display on-horizon divergences.

First there are cases where at least one of the NP variables vanishes (others nonzero):
\begin{align}
&\text{Case 0A:}&&\phi_0=\phi_2=0\\
&\text{Case 0B:}&&\phi_2=0\quad(\text{or }\phi_0=0)\\
&\text{Case 0C:}&&\phi_1=0.
\end{align}
More general cases where all NP variables are nonzero, subject to the degeneracy condition $\Im(\phi_1^2)=\Im(\phi_0\phi_2)=-\Im(\Phi_2^2)/(8\Delta_r)$ (noting $\Im\Phi_1=0$ for SAS solutions), can be organized as follows
\begin{align}
\text{for }\Im(\phi_1^2)=0,&\notag\\
&\text{Case 1A:}&&\Re\phi_1=0\\
&\text{Case 1B:}&&\Im\phi_1=0\\
\text{for }\Im(\phi_1^2)\neq0,&\notag\\
&\text{Case 2A:}&&\Re(\phi_1^2)=-\frac{\Re(\Phi_2^2)}{8\Delta_r}(=0\text{ for a subcase})\\
&\text{Case 2B:}&&\Re(\phi_1^2)=0,\;-\frac{\Re(\Phi_2^2)}{8\Delta_r}\neq0\\
&\text{Case 2C:}&&\Re(\phi_1^2)\neq0,\;-\frac{\Re(\Phi_2^2)}{8\Delta_r}=0\\
&\text{Case 3:}&&\Re(\phi_1^2)\neq-\frac{\Re(\Phi_2^2)}{8\Delta_r}\neq0.
\end{align}
We find solutions for Case 0C and show that (in Appendix) there are no solutions for Cases 0A, 0B, 1A, 1B and 2A. (There may be better ways to categorize Cases 2A--3 taking into account $\Phi_1$.)

\subsection{SAS solutions for Case 0C: $\phi_1=0$}\label{case1}

The force-free equations \eqref{FJ_eqns} reduce to (the first two being equivalent)
\begin{equation}\label{FJ_zero_phi1}
\Re(\bar\phi_0J_m)=0,\qquad\bar\phi_0J_n-\phi_2J_l=0,
\end{equation}
which imply
\begin{equation}\label{c}
\frac{J_n}{J_l}=c(r,\theta,\varphi,t)=\frac{\phi_2}{\bar\phi_0}
\end{equation}
for some real function $c(r,\theta,\varphi,t)$. Note that
\begin{equation}\label{phi2'=c*phi1'}
\frac{\phi_2'}{\bar\phi_0'}=c(r,\theta,\varphi,t)\frac{\Sigma}{\Delta_r}.
\end{equation}
In addition, one needs to check the conditions \eqref{dF=0_NP_J} to ensure the homogeneous Maxwell equations. The electromagnetic field is degenerate and non-null, i.e., $\phi_0\phi_2-\phi_1^2\propto I_1+iI_2$ is real and non-zero. Explicitly, we have for the two force-free equations \eqref{FJ_zero_phi1} and the two conditions from \eqref{dF=0_NP_J}
\begin{align}
\bar J_n=J_n=cJ_l\qquad&\Rightarrow\qquad\nabla_{\bar m}(\phi'_0\bar\phi'_2)=0\label{J_nl}\\
\Re(\bar\phi_0J_m)=0\qquad&\Rightarrow\qquad2\Re(\bar\phi_0'\nabla_n\phi_0')=\nabla_n(\bar\phi_0'\phi_0')=0\label{D[n]phi0'}\\
\bar{J}_{\bar{m}}=J_m\qquad&\Rightarrow\qquad\frac{\Sigma}{\Delta_r}\nabla_n\phi_0'+\nabla_l\bar\phi_2'=0\label{J_mbarm}\\
J_n=\bar{J}_n\qquad&\Rightarrow\qquad\sin\theta\Delta_\theta\partial_\theta\Im\phi_2'+(\Xi\partial_\varphi+a\sin^2\theta\partial_t)\Re\phi_2'=0,\label{J_n_real}
\end{align}
where we have eliminated $c$ using (the conjugate of) \eqref{phi2'=c*phi1'}.

We will mainly concentrate on SAS solutions in the sense that $F_{\varphi t}\propto\Im\Phi_1=0$, which implies, by \eqref{Phi1Phi2}, \eqref{phi'} and \eqref{phi2'=c*phi1'},
\begin{equation}
\Im\bigl[\phi'_0+\frac{2\Sigma}{\Delta_r}c(r,\theta,\varphi,t)\bar\phi_0'\bigr]=0,
\end{equation}
which breaks up into two subcases
\begin{equation}\label{2cases_SAS}
\Im\phi'_0=0\qquad\text{and}\qquad c(r,\theta,\varphi,t)=\frac{\Delta_r}{2\Sigma}.
\end{equation}

\subsubsection{Subcase 0C-1: SAS solutions with $\Im\phi'_0=0$}\label{case1.1}

In this subcase we additionally have $\Im\phi'_2=0$ by \eqref{phi2'=c*phi1'}. Then \eqref{J_nl}, \eqref{D[n]phi0'}, \eqref{J_mbarm} and \eqref{J_n_real} reduce respectively to (keeping $\partial_\varphi,\partial_t$ terms for now)
\begin{align}
\Re\eqref{J_nl}\qquad\Leftrightarrow\qquad&\partial_\theta(\phi'_0\phi'_2)=0\label{J_nl1_Re}\\
\Im\eqref{J_nl}\qquad\Leftrightarrow\qquad&(\Xi\partial_\varphi+a\sin^2\theta\partial_t)(\phi'_0\phi'_2)=0\label{J_nl1_Im}\\
&\nabla_n\phi_0'=0\label{D[n]phi0'1}\\
&\nabla_l\phi'_2=0\label{J_mbarm1}\\
&(\Xi\partial_\varphi+a\sin^2\theta\partial_t)\phi_2'=0\label{J_n_real1}.
\end{align}
It is easy to deduce from \eqref{J_nl1_Re} and \eqref{J_nl1_Im} that
\begin{equation}\label{phi0sq_c=pr}
\phi'_0\phi'_2=p(r),
\end{equation}
for some function $p(r)$. We again transform to the ingoing Kerr coordinates $\{v,r,\theta,\psi\}$ as in \cref{sec_Jac}. Then \eqref{D[n]phi0'1} is simply $\partial_r\phi_0'=0$ and \eqref{J_mbarm1} and \eqref{J_n_real1} become, using \eqref{phi0sq_c=pr},
\begin{align}
{\Delta_r\partial_r\ln p(r)-2[a\Xi\partial_\psi+(r^2+a^2)\partial_v]}\phi_0'(v,\theta,\psi)&=0\\
(\Xi\partial_\psi+a\sin^2\theta\partial_v)\phi_0'(v,\theta,\psi)&=0.
\end{align}
The solution formally takes the form
\begin{equation}\label{phi0'qp}
\phi_0'(v,\theta,\psi)=q(\theta)\exp\Bigl[\frac{\Delta_r\partial_r\ln p(r)}{\Sigma}\big(v-\frac{a\sin^2\theta}{\Xi}\psi\bigr)\Bigr],
\end{equation}
where $q(\theta)$ is some function. For the r.h.s to be $r$-independent, one must have $p(r)=p=\text{const.}$, 
and the solution is indeed ($\psi,v$)-independent:
\begin{equation}\label{case1_sln_Phi}
\phi_0'=q(\theta),\qquad\phi_2'=\frac{p}{q(\theta)},\qquad\Phi_{1,2}=\frac{q(\theta)\pm2\frac{p}{q(\theta)}}{\sqrt{\Delta_\theta}\sin\theta}.
\end{equation}
Unfortunately, the solution is singular on the horizon $\Delta_r=0$, for by \eqref{F_rth_IK}
\begin{equation}
F_{r\theta}^\mathrm{IK}=-\frac{2\sqrt2\Sigma}{\Delta_r\Delta_\theta\sin\theta}\frac{p}{q(\theta)},
\end{equation}
and due to the earlier constraint $p(r)=p$ the divergence cannot be removed. In fact, when $p=0$ we recover the null current solution \eqref{S(theta)}, with $q(\theta)\mapsto\Delta_\theta\sin\theta S'(\theta)/\Xi$. $p$ determines the causal nature of the current: $J_{(a)}J^{(a)}=2J_lJ_n=p[\partial_\theta\ln q(\theta)]^2/(\Delta_r\Sigma\sin^2\theta)$, which indeed diverges on the horizon unless $p=0$ ($\partial_\theta q(\theta)=0$ just gives a vanishing current).

\paragraph*{Non-rotating limit}

For Schwarzschild-AdS black holes we can remove the horizon divergence. Requiring $[\Delta_r\partial_r\ln p(r)/\Sigma]\rvert_{a=0}=\text{const.}$ in \eqref{phi0'qp} yields 
\begin{equation}
p(r)=C(r^2+rr_H+r_H^2+l^2)^{-\frac{k}{2}}(r-r_H)^k\exp\Bigl[k\frac{3r_H^2+2l^2}{r_H\sqrt{3r_H^2+4l^2}}\arctan\Bigl(\frac{2r+r_H}{\sqrt{3r_H^2+4l^2}}\Bigr)\Bigr],
\end{equation}
where $C$ is constant, $r_H$ is the horizon and we have fixed the `const.' so that $p(r)\propto(r-r_H)^k$. $F_{r\theta}^\mathrm{IK}$ is now regular both on the horizon and at infinity. It is however worth noting the Schwarzschild limit (setting $l\rightarrow\infty$ before solving for $p(r)$) leads to $p(r)=C(r-r_H)^ke^{kr/r_H}$ which diverges at large $r$.

\subsubsection{Subcase 0C-2: SAS solutions with $c(r,\theta,\varphi,t)=\Delta_r/(2\Sigma)$}\label{case2}

We now turn to the second subcase in \eqref{2cases_SAS}, which implies
\begin{equation}
\phi'_2=\frac{\bar\phi'_0}{2}.
\end{equation}
We rewrite, again in the ingoing Kerr coordinates,
\begin{equation}
\phi'_0=\exp[R(r,\theta,\psi,v)+i\Theta(r,\theta,\psi,v)],\qquad\phi'_2=\frac{\exp[2R(r,\theta,\psi,v)]}{2\phi'_0},
\end{equation}
for arbitrary functions $R$ and $\Theta$. Then \eqref{J_nl}, \eqref{D[n]phi0'} and \eqref{J_mbarm} reduce respectively to
\begin{align}
\Re\eqref{J_nl}\qquad&\Leftrightarrow\qquad\sin\theta\Delta_\theta\partial_\theta R+a\sin^2\theta\partial_v\Theta-\Xi\partial_\psi\Theta=0\label{J_nl2_Re}\\
\Im\eqref{J_nl}\qquad&\Leftrightarrow\qquad\sin\theta\Delta_\theta\partial_\theta\Theta+a\sin^2\theta\partial_vR-\Xi\partial_\psi R=0\label{J_nl2_Im}\\
\eqref{D[n]phi0'}\qquad&\Leftrightarrow\qquad\partial_rR=0\label{partial_rR=0}\\
\Re\eqref{J_mbarm}\qquad&\Leftrightarrow\qquad(r^2+a^2)\partial_vR+a\Xi\partial_\psi R=0\label{J_mbarm2_Re}\\
\Im\eqref{J_mbarm}\qquad&\Leftrightarrow\qquad(r^2+a^2)\partial_v\Theta+a\Xi\partial_\psi\Theta=0\label{J_mbarm2_Im},
\end{align}
and \eqref{J_n_real} turns out to be equivalent to \eqref{J_nl}. \eqref{partial_rR=0} and \eqref{J_mbarm2_Re} imply $R=R(\theta)$, and then \eqref{J_nl2_Im} implies $\Theta=\Theta(r,\psi,v)$. The remaining two equations \eqref{J_nl2_Re} and \eqref{J_mbarm2_Im} give the solution
\begin{equation}
R(\theta)=R=\text{const.},\qquad\Theta(r,\psi,v)=\Theta(r).
\end{equation}
The regularity conditions in \cref{sec_ff} require $\phi_2'\propto\exp[i\Theta(r)]$ to vanish on the horizon which is impossible. The current is however finite: $J_{(a)}J^{(a)}=-2J_mJ_{\bar m}=-e^{2R}[\partial_r\Theta(r)]^2/(2\Sigma\Delta_\theta\sin^2\theta)$. (A spacelike current is perfectly physical in the presence of both positive and negative charges, in particular for magnetically dominant configurations).

\subsection{Further possible special SAS solutions}\label{othercases}

The above cases have exhausted the possibilities with $\Im(\phi_1^2)=0$. For Cases 2A--C in \cref{classes} we first rewrite
\begin{equation}
\Phi_2+i\sqrt{8\Delta_r}\phi_1\equiv R_1e^{i\Theta},\qquad \Phi_2-i\sqrt{8\Delta_r}\phi_1\equiv R_2e^{-i\Theta}
\end{equation}
for real functions $R_{1,2},\Theta$, so that the degeneracy condition becomes $\Im(R_1R_2)$=0 which always holds. Then the cases are equivalently characterized as
\begin{align}
&\text{Case 2A:}&&R_1R_2=0,\qquad(\text{subcase }\Re(\Phi_2^2)=0=\Re(\phi_1^2):\;\tan\Theta=\pm1)\\
&\text{Case 2B:}&&R_1R_2\neq0,\qquad\tan\Theta=\pm\frac{R_1-R_2}{R_1+R_2}\\
&\text{Case 2C:}&&R_1R_2\neq0,\qquad\tan\Theta=\pm\frac{R_1+R_2}{R_1-R_2}\\
&\text{Case 3:}&&R_1R_2\neq0.
\end{align}
Note that to exclude previous cases (with $\Im(\phi_1^2)=0$) one has $R_1\neq\pm R_2$. We show in the appendix that Case 2A has no solutions in the $a=0$ limit, implying there are no solutions for nonzero $a$ too. Future work would include exploring the remaining cases.

\section{Duality rotation}\label{sec_rot}

\subsection{Definition and applications to the force-free problem}

The above solutions are special in the sense that one imposes $\phi_{1,2}=0$ or $\phi_1=0$. In this section we explore possibilities of reinterpreting these solutions via duality rotations. A \emph{duality rotation} is defined by the following operation on the field tensor
\begin{equation}\label{duality_rotation1}
F\rightarrow\tilde F\equiv F\cos\alpha+{^\star}\!F\sin\alpha,
\end{equation}
where the rotation angle $\alpha$ is a spacetime function. Equivalently,
\begin{equation}\label{duality_rotation2}
\tilde F^-=e^{-i\alpha}F^-,\qquad\tilde\phi_{0,1,2}=e^{-i\alpha}\phi_{0,1,2},
\end{equation}
where $F^-\equiv(F+i\,{^\star}\!F)/2$ is the anti-self-dual part of $F$ and the second expression is because the NP variables are expansion coefficients of $F^-$ in the anti-self-dual bivector basis \footnote{Namely, $F^-_{\mu\nu}=\phi_0U_{\mu\nu}+\phi_1W_{\mu\nu}+\phi_2V_{\mu\nu}$, with $U_{\alpha\beta}\equiv 2\bar m_{[\alpha}n_{\beta]},W_{\alpha\beta}\equiv 2(n_{[\alpha}l_{\beta]}+m_{[\alpha}\bar m_{\beta]}),V_{\alpha\beta}\equiv 2l_{[\alpha}m_{\beta]}$.}.

Duality rotation has the merit that it leaves the electromagnetic energy-momentum tensor invariant and solutions to the conservation equations \eqref{ff} come in pairs $\{F_{\mu\nu},\tilde F_{\mu\nu}\}$. In particular, it can be used to construct force-free solutions from vacuum solutions which satisfy the conservation equations trivially. It is also interesting to check if an existing force-free solution admits a ``dual'' vacuum description. In the following, we save `$\tilde F$' for the vacuum solution and `$F$' for the force-free solution, related as in \eqref{duality_rotation1} in a duality pair (if it exists).

The conditions of existence of duality pairs can be inferred by noting that, under a duality rotation,
\begin{align}
\tilde I&=e^{-2i\alpha}I\\
\tilde J^\mu+i\tilde L^\mu&=e^{-i\alpha}\bigl[(J^\mu+iL^\mu)-i{F^-}^{\mu\nu}\alpha_{,\nu}\bigr],\qquad(L\equiv{^\star}\!F^{\mu\nu}_{;\nu}),\label{JL_rot}
\end{align}
for the invariants and currents. We require $\tilde L=L=0$ and rewrite \eqref{JL_rot} (inversely through changing $\tilde J\leftrightarrow J,\alpha\rightarrow-\alpha$ and setting $\tilde J=0$) as
\begin{equation}\label{J_ff}
J^\mu=ie^{i\alpha}({\tilde F}^-)^{\mu\nu}\alpha_{,\nu},
\end{equation}
which can be used in the case where one duality-rotates a vacuum solution $\tilde F$ to a force-free solution with current $J$. In this case, the rotation angle is guaranteed to exist, given by (using $I_2=0$)
\begin{equation}\label{tan2alpha}
\tan2\alpha=-\frac{\tilde I_2}{\tilde I_1},
\end{equation}
but one needs to additionally check that the current \eqref{J_ff} is indeed real (i.e.\ $L=0$). For the other case, where one wants to duality-rotate a known force-free solution to a vacuum one, the rotation angle is given by \cite{Geometrodynamics}

\begin{equation}\label{alpha_mu}
\alpha_\mu\equiv\alpha_{,\mu}=\frac{2}{I_1}F_{\nu\mu}J^\nu,
\end{equation}
subject to the integrability condition. We next consider examples from both cases.

\subsection{Duality rotation for null current configuration}\label{rot_null}

Assume that the configuration on the force-free side is that of Brennan et al. as in \cref{sec_Jac}, and we look for a duality rotation such that
\begin{align}
\text{force-free}&\begin{cases}
\phi_0\neq0,\phi_1=\phi_2=0\\
J_l=\mathcal{J},J_{n,m,\bar m}=0\\
I_1=I_2=0
\end{cases} & &\overset{e^{-i\alpha}}{\longrightarrow} &
\text{vacuum}&\begin{cases}
\tilde\phi_0\neq0,\tilde\phi_1=\tilde\phi_2=0\\
\tilde J_{(a)}=0\\
\tilde I_1=\tilde I_2=0,
\end{cases}\label{Jac_rot}
\end{align}
The real field quantities on both sides are given by
\begin{align}
B_T+iF_{\varphi t}&=\frac{\sin\theta\sqrt{\Delta_\theta}}{\sqrt2\Xi}\Phi_1,\\
F_{rt}=-\frac{a\Xi}{r^2+a^2}F_{r\varphi}&=\frac{a\sin\theta\sqrt{\Delta_\theta}}{\sqrt2\Delta_r}\Im\Phi_1,\\
F_{\theta t}=-\frac{\Xi}{a\sin^2\theta}F_{\theta\varphi}&=\frac{\Re\Phi_1}{\sqrt2\sqrt{\Delta_\theta}},
\end{align}
and similarly for tilde quantities.

\subsubsection{force-free $\mapsto$ vacuum}

As a first example, we duality-rotate the known SAS force-free solution \eqref{S(theta)} to a vacuum one. We evaluate the vacuum equations (r.h.s.\ of \eqref{Jac_rot}), replacing $\tilde\phi_0\rightarrow e^{-i\alpha}\phi_0$ in $\tilde J$ where $\phi_0$ is the force-free solution \eqref{S(theta)}. Then we find the following first-order differential equations for $\alpha$ (equivalent to \eqref{alpha_mu} which however is not applicable for $I=0$):
\begin{align}
0=&\Im\tilde J_l=\frac{\Re\Phi_1}{\sqrt2\Delta_r}\Bigl[\frac{\sin\alpha}{\sqrt{\Delta_\theta}\sin\theta}[\Xi\alpha_\psi+a\sin^2\theta\alpha_v]-\sqrt{\Delta_\theta}\cos\alpha\alpha_\theta\Bigr]-\sin\alpha\mathcal{J}(r,\theta)\label{Im_J_rot_SAS}\\
0=&\Re\tilde J_l=\Im\tilde J_l(\sin\alpha\rightarrow-\cos\alpha,\cos\alpha\rightarrow\sin\alpha),\label{Re_J_rot_SAS}\\
0=&\tilde J_m\Rightarrow\nabla_n\alpha,
\end{align}
with $\mathcal{J}$ and $\Re\Phi_1$ related through \eqref{S(theta)}. Use again the ingoing Kerr coordinates, and the last equation is just $\partial_r\alpha=0$. Canceling $\mathcal{J}$ using the first two equations, one has $\alpha_\theta=0$. Then the $(\psi,v)$-independence of the whole equation \eqref{Im_J_rot_SAS} imposes $\alpha=c_0+c_1\psi+al^{-2}c_2v$ for constant $c_{0,1,2}$. For such $\alpha$, \eqref{Im_J_rot_SAS} and \eqref{S(theta)} yield
\begin{align}
\Re\Phi_1&=C\Bigl(\frac{l-a\cos\theta}{l+a\cos\theta}\Bigr)^{\frac{a(c_2-c_1)}{2l}}(1-\cos\theta)^{\frac{c_1-1}{2}}(1+\cos\theta)^{-\frac{c_1+1}{2}}\\
\mathcal{J}&=\frac{\Re\Phi_1}{\sqrt2\Delta_r\sqrt{\Delta_\theta}\sin\theta}(c_1\Xi+c_2a^2l^{-2}\sin^2\theta),
\end{align}
which is the concrete force-free configuration that admits a vacuum dual.

To check the regularity conditions in \cref{reg_conditions}, we compute components of the current in the Cartesian Kerr-Schild coordinates as
\begin{equation}\label{J_xyz_1}
J^\tau=\frac{\Delta_r}{2\Sigma}\mathcal{J},\qquad J^z=-\cos\theta\frac{\Delta_r}{2\Sigma}\mathcal{J},\qquad J^x=-\cos\psi\sin\theta\frac{\Delta_r}{2\Sigma}\mathcal{J},\qquad J^y=-\sin\psi\sin\theta\frac{\Delta_r}{2\Sigma}\mathcal{J},
\end{equation}
which are all regular on the horizon. At the poles, one has $\mathcal{J}\sim\mathcal{O}(\sin^{-3}\theta)$ if $c_1\neq0$ and $\mathcal{J}\sim\mathcal{O}(1)$ if $c_1=0$; for the latter, the above components are all regular. The field itself however fails to meet the criterion in \eqref{poles_NP} ($\Re\Phi_1\sim\mathcal{O}(\sin^{-1}\theta)$ for $c_1=0$), which could be associated with the finite charge and current densities $J^\tau,J^z$ as shown in \eqref{J_xyz_1}. The vacuum solution is wavelike:
\begin{equation}
\tilde\Phi_1=\Re\Phi_1e^{-i(c_0+c_1\psi+al^{-2}c_2v)},
\end{equation}
also scaling as $\mathcal{O}(\sin^{-1}\theta)$. (See \cite{Jacobson} for discussions on non-existence of globally regular null vacuum solutions.)

[For the special case $\alpha=\alpha(\theta)$, solving \eqref{Im_J_rot_SAS} for $\alpha$ first and evaluating \eqref{Re_J_rot_SAS} yields $\csc\alpha=C\Xi\sqrt{\Delta_\theta}\sin\theta\Re\Phi_1,\;\Re\tilde J_l=\csc\alpha\mathcal{J}$, i.e., not a vacuum solution but a rescaling of the original one.]

\subsubsection{vacuum $\mapsto$ force-free}

As a second example, we instead solve the vacuum equations explicitly first and find appropriate $\alpha$. For simplicity concentrate on the Kerr limit. The vacuum equations are formally the same as \eqref{Jl_eqn_Re} \& \eqref{Jl_eqn_Im}, treating the quantities there as the tilde ones and setting $\tilde{\mathcal J}=0$. The general solution is $\tilde\Phi_1=X\bigl[\psi+i\ln(\csc\theta-\cot\theta),-v+ia\cos\theta\bigr]\csc\theta$ for an arbitrary function $X(,)$. We fix $X$ so that
\begin{equation}
\tilde\Phi_1(\theta,\psi,v)=e^{w_1[\psi+i\ln(\csc\theta-\cot\theta)]-w_2(v-ia\cos\theta)}\csc\theta,
\end{equation}
where $w_1,w_2$ are constants. The current on the force-free solution side is directly calculated as $\mathcal{J}=J_l(\Phi_1=e^{i\alpha}\tilde\Phi_1)$, whose reality condition we need to solve for $\alpha$ (cf.\ \eqref{J_ff}).

For real $w_1,w_2$, we look at a special case $\alpha=\alpha(\theta)$ and find
\begin{equation}
\alpha(\theta)=\frac{\pi}{2}-w_1\ln(\csc\theta-\cot\theta)-w_2a\cos\theta.
\end{equation}
The force-free solution is then
\begin{equation}
\Phi_1=ie^{w_1\psi-w_2v}\csc\theta,\qquad\mathcal{J}=\sqrt2e^{w_1\psi-w_2v}(w_1\csc\theta-w_2a\sin\theta).
\end{equation}
There are no horizon divergences; at the poles, we note that
\begin{equation}\label{J_xyz_2}
J^\tau=\csc\theta\frac{\mathcal{J}}{4\Sigma},\qquad J^z=-\cot\theta\frac{\mathcal{J}}{4\Sigma},\qquad J^x=-\cos\psi\frac{\mathcal{J}}{4\Sigma},\qquad J^y=-\sin\psi\frac{\mathcal{J}}{4\Sigma},
\end{equation}
so for them to be regular one needs to set $w_1=0$. The field is always singular: $\Phi_1\sim\mathcal{O}(\sin^{-1}\theta)$.

For imaginary $w_1=i\hat w_1,\,w_2=i\hat w_2$, we also obtain a special solution
\begin{align}
\alpha&=\alpha_1(\theta)-\hat w_1\psi+\hat w_2v,\qquad\text{with }\sin[\alpha_1(\theta)]=Ce^{\hat w_1\ln(\csc\theta-\cot\theta)+\hat w_2a\cos\theta}\\
\Phi_1&=Ce^{i\alpha_1(\theta)}\csc[\alpha_1(\theta)]\csc\theta,\qquad\mathcal{J}=-2\sqrt2C\csc[2\alpha_1(\theta)](\hat w_1\csc\theta-\hat w_2a\sin\theta),
\end{align}
where $C$ is constant. In fact for the correct range of $\sin\alpha_1$ one has $\hat w_1=0$ and $\lvert C\rvert\leq e^{-\lvert\hat w_2\rvert a}$. Then the current components (same forms as in \eqref{J_xyz_2}) are regular at the poles, but the field $\Phi_1$ is again singular.

\subsection{Duality rotation for non-null current configuration with $\phi_1=0$}

We now look for a duality pair for the configuration Case 0C: $\phi_1=0=\tilde\phi_1$. Directly applying the formula \eqref{alpha_mu} to the SAS force-free solutions in \cref{case1} shows that appropriate $\alpha$ does not exist. We thus proceed by assuming a more general situation where such $\alpha$ does exist. Here both solutions in the duality pair are unknown and we need to manipulate available equations altogether.

First, evaluate the conditions \eqref{J_mbarm} \& \eqref{J_n_real} from the force-free side, replacing $\phi'_{0,2}\rightarrow e^{i\alpha}\tilde\phi'_{0,2}$, and use the fact that $\tilde\phi'_{0,2}$ solves the vacuum equations, to get
\begin{align}
\bar J_{\bar m}=J_m\qquad&\Rightarrow\qquad(1+c_1)\Delta_r\alpha_r+(1-c_1)\bigl[a\Xi\alpha_\varphi+(r^2+a^2)\alpha_t\bigr]=0\label{J_mbarm_b}\\
J_n=\bar J_n\qquad&\Rightarrow\qquad\Delta_\theta\sin\theta\alpha_\theta-\frac{\Im\phi'_2}{\Re\phi'_2}\bigl[\Xi\alpha_\varphi+a\sin^2\theta\alpha_t\bigr]=0,\label{J_n_real_b}
\end{align}
where $c_1=2c\Sigma/\Delta_r$ with $c$ defined in \eqref{phi2'=c*phi1'}. To solve for $\alpha$, we assume $\alpha_\theta=0$, which implies $\alpha_\varphi=\alpha_t=0$. So $\alpha=\alpha(r),\,c_1=-1$ from \eqref{J_mbarm_b}. (Note that solutions in \cref{case2} have $c_1=1$.)

Now construct force-free solutions with $c_1=-1\,\Leftrightarrow\,\phi_2'=-\bar\phi'_0/2$. It is straightforward to derive from \eqref{J_nl}--\eqref{J_n_real} that
\begin{align}
\phi'_0&=C\Bigl(\frac{l-a\cos\theta}{l+a\cos\theta}\Bigr)^{\frac{a(w_1-w_2)}{2l}}(1-\cos\theta)^{-\frac{w_1}{2}}(1+\cos\theta)^{\frac{w_1}{2}}e^{i(w_0+w_1\varphi+al^{-2}w_2t)}\label{phi0_case3}\\
J_m&=-ia\frac{\phi'_0}{2\rho\Sigma\Delta_r\sqrt{\Delta_\theta}\sin\theta}\bigl[\Xi w_1+l^{-2}(r^2+a^2)w_2\bigr],\qquad J_l=J_n=0,
\end{align}
with constant $C,w_{0,1,2}$. While the field can be regular (for $w_1=0$), the current displays singularities both on the horizon and at the poles:
\begin{align}
J^\tau&=J^v,\qquad J^z=-r\sin\theta J^\theta\\
J^x&=(r\cos\psi-a\sin\psi)\cos\theta J^\theta-(r\sin\psi+a\cos\psi)\sin\theta J^\psi\\
J^y&=(r\sin\psi+a\cos\psi)\cos\theta J^\theta+(r\cos\psi-a\sin\psi)\sin\theta J^\psi,
\end{align}
where
\begin{equation}
J^\theta=\sqrt{2\Delta_\theta}\Re(\rho J_m),\qquad J^v=J^\psi=\frac{\sqrt2\Xi}{\sqrt{\Delta_\theta}\sin\theta}\Im(\rho J_m).
\end{equation}

Finally, on the vacuum side, write the solution as $\tilde\phi'_{0,2}=e^{-i\alpha(r)}\phi'_{0,2}$ in terms of the above force-free solution, and evaluate the current $\tilde J$ to find further constraints on $\alpha$:
\begin{equation}
\tilde J_{(a)}=0\qquad\Rightarrow\qquad\alpha_r=-\frac{a\Xi}{\Delta_r}w_1-\frac{r^2+a^2}{\Delta_r}w_2\frac{a}{l^2}.
\end{equation}
Note that the factors in front of $w_1,w_2$ appear in the transformation rules \eqref{ingoingKerr} for the ingoing Kerr coordinates.

\section{Other special SAS solutions}\label{sec_SAS}

\subsection{General force-free equations}

We return to the normal formalism of the force-free magnetospheres and concentrate on the stationary and axisymmetric case with $F_{\varphi t}=0$. Then a degenerate $F_{\mu\nu}$ can be specified by $A_{\varphi,r}$, $A_{\varphi,\theta}$, $B_T$ \& $\omega$ \footnote{Alternative quantities often used in the literature (e.g.\ \cite{ZhangLehner:2014ff}) are $\psi$ (stream function), $I$ (polar current) and $\Omega_F$ (angular velocity of field lines), corresponding to $A_\varphi$, $B_T$ and $\omega$ respectively.}. Defining the bracket notation $\{X,Y\}\equiv X_{,r}Y_{,\theta}-Y_{,r}X_{,\theta}$ for any functions $X$ and $Y$, one has by the definition of $\omega$
\begin{equation}\label{omega(A_varphi)}
\{A_\varphi,\omega\}=0\quad\Rightarrow\quad\omega=\omega(A_\varphi).
\end{equation}

The degeneracy condition also implies that only two of the four force-free equations \eqref{ff} are independent (since an antisymmetric $F_{\mu\nu}$ with $\det F_{\mu\nu}=0$ has rank 2). Indeed, the ($\varphi,t$)-components of \eqref{ff} are proportional to each other:
\begin{equation}\label{FJ_pht}
F_{\mu\varphi}J^\mu=-\frac{1}{\omega}F_{\mu t}J^\mu=A_{\varphi,r}J^r+A_{\varphi,\theta}J^\theta=0,\qquad\text{with } J^r=-\frac{(B_T)_{,\theta}}{\sqrt{-g}},\quad J^\theta=\frac{(B_T)_{,r}}{\sqrt{-g}},
\end{equation}
implying
\begin{equation}
\{A_\varphi,B_T\}=0\quad\Rightarrow\quad B_T=B_T(A_\varphi).
\end{equation}
Same for the ($r,\theta$)-components, using \eqref{FJ_pht}:
\begin{equation}\label{FJ_rth}
\frac{1}{A_{\varphi,r}}F_{\mu r}J^\mu=\frac{1}{A_{\varphi,\theta}}F_{\mu\theta}J^\mu=\frac{[B_T^2(A_\varphi)]'}{2h_{\varphi\varphi}\alpha^2}-J_\varphi+\omega J_t=0,
\end{equation}
where a prime denotes derivative w.r.t.\ the argument. Here note that second derivatives of $A_\varphi$ only appear in $\omega J_t-J_\varphi$ in \eqref{FJ_rth}, which is the main equation to solve in the force-free problem. An equivalent form of \eqref{FJ_rth} is
\begin{equation}\label{dT_rth}
\frac{1}{\sqrt{-g}}\bigl(c_\omega\sqrt{-g}h^{MN}A_{\varphi,N}\bigr)_{,M}+\frac{(h^{MN}A_{\varphi,M}A_{\varphi,N})h_{\varphi\varphi}(\beta^\varphi+\omega)\omega'+\frac{1}{2}(B_T^2)'}{h_{\varphi\varphi}\alpha^2}=0,
\end{equation}
where $M,N=r,\theta$, $c_\omega\equiv h_{\varphi\varphi}^{-1}-(\beta^\varphi+\omega)^2\alpha^{-2}$, and $\omega$ \& $B_T$ are viewed as functions of $A_\varphi$ \footnote{We note that the first term in \eqref{dT_rth} can be written in a slightly more concise form as $c_\omega^2\bar\Box A_\varphi$, while the second term gets multiplied by $c_\omega^2$, where $\bar\Box$ is the 4-D d'Alembertian associated with the Weyl transformed metric $\bar g_{\mu\nu}=c_\omega g_{\mu\nu}$ (noting $\partial_t,\partial_\varphi\rightarrow0$).}.
The quadratic terms $A_{\varphi,r}^2$ \& $A_{\varphi,\theta}^2$ in \eqref{dT_rth} can be made linear using $\omega'=\omega_{,r}/A_{\varphi,r}=\omega_{,\theta}/A_{\varphi,\theta}$. It is in fact possible to completely remove the second-order derivatives of $A_\varphi$ by choosing appropriate $\omega$, which we explore in the next subsection.

\subsection{Reduction to first-order equation: solutions in flat/pure AdS spacetimes}

The second derivatives of $A_\varphi$ (and thus the whole first term) in \eqref{dT_rth} vanish if $c_\omega=0$, i.e.,
\begin{equation}\label{Omega+-}
\omega=\Omega^\pm\equiv-\beta^\varphi\pm\frac{\alpha}{\sqrt{h_{\varphi\varphi}}},
\end{equation}
where the functions $\Omega^\pm$ appear in the constraint $\Omega^-<\Omega<\Omega^+$ for the Killing vector $\xi_{(t)}^\mu+\Omega\xi_{(\varphi)}^\mu$ to be timelike.
Then equation \eqref{dT_rth} becomes
\begin{equation}\label{ff_1st}
h^{rr}A_{\varphi,r}^2+h^{\theta\theta}A_{\varphi,\theta}^2\pm\frac{[B_T^2(A_\varphi)]'}{2\alpha\sqrt{h_{\varphi\varphi}}\omega'(A_\varphi)}=0,
\end{equation}
where we have restricted to the case of nonzero $\omega'(A_\varphi)$, i.e., invertible $\omega(A_\varphi)$, which we need for the following derivations. We next construct some solutions for the flat and pure AdS backgrounds.

\subsubsection{flat spacetime}
The flat limit is obtained by setting $m=a=0$ and $l\rightarrow\infty$, for which
\begin{equation}
\omega_\text{flat}=\pm\frac{1}{r\sin\theta}.
\end{equation}
Introducing new coordinates $\{x\equiv r\sin\theta,\,z\equiv r\cos\theta\}$ (which are radial and axial directions of cylindrical coordinates), we have $\omega_\text{flat}=\pm1/x$ and the bracket condition \eqref{omega(A_varphi)} yields $A_{\varphi,z}=0$. Then \eqref{ff_1st} becomes
\begin{equation}\label{ff_1st_x}
A_{\varphi,x}^2+\frac{[B_T^2(A_\varphi)]'}{2x\omega_\text{flat}'(A_\varphi)}=A_{\varphi,x}^2\mp\frac{x}{2}[B_T^2(x)]'=0.
\end{equation}
Solutions can be found by specifying $B_T(x)$ or $B_T(A_\varphi)$ \footnote{A solution $A_\varphi(x)$ or relation $B_T(A_\varphi)$ partially fixes the gauge (where $x$ and $B_T$ are gauge invariant). Generally, requiring that the $\varphi$- and $t$-independence of the vector potential be preserved under a gauge transformation $A_\mu(r,\theta)\rightarrow A_\mu(r,\theta)+\partial_\mu\alpha(r,\theta,\varphi,t)$, one finds $\alpha(r,\theta,\varphi,t)=\alpha_1\varphi+\alpha_2 t+\alpha_3(r,\theta)$ for constant $\alpha_1,\alpha_2$ and arbitrary $\alpha_3(r,\theta)$. Then $A_\varphi\rightarrow A_\varphi+\alpha_1$, i.e., a constant shift.}. Note that $B_T=-x^2B^\varphi$. $\omega$ is divergent on the axis, and we may expect more or less singular behavior of the current and field components. The quantity that we want to make regular is $\sqrt{-g}J^\varphi$, which should be $\sim\mathcal{O}(x)$ or higher, where note that $\sqrt{-g}=x$. 
A sample solution to \eqref{ff_1st_x} with the plus sign in front of the second term is given by
\begin{equation}
B_T^2(A_\varphi)=\begin{cases}
cA_\varphi^n,\quad(n\neq2)\\
cA_\varphi^2
\end{cases},\qquad A_\varphi(x)=\begin{cases}
\bigl[\frac{1}{4}cn(n-2)x^2+C\bigr]^{\frac{1}{2-n}},\quad(n\neq2)\\
Ce^{-\frac{cx^2}{2}}
\end{cases}.
\end{equation}
On the axis $x=0$, we have a vanishing toroidal current density $\sqrt{-g}J^\varphi\sim\mathcal{O}(x)$ but finite axial current and charge densities $\sqrt{-g}J^z,\sqrt{-g}J^t\sim\mathcal{O}(1)$. $J^x$ vanishes identically. More regular solutions are also possible, e.g., $\sqrt{-g}J^\varphi\sim\mathcal{O}(x^3)$ for $B_T^2=\arctan x^2$ with the minus sign in \eqref{ff_1st_x} \footnote{For solutions obtained by specifying $B_T^2(x)$, a shift $B_T^2(x)\rightarrow B_T^2(x)+\text{const.}$ does not change $A_\varphi$ but can change the leading order of $B_T$ and $J^z$ at $x=0$ in either direction. E.g., a more regular $B_T$ does not have to be associated with a more regular $J^z\sim B_T'(x)$.}. Asymptotically, the currents are (or can be made) regular. The non-vanishing poloidal energy and angular momentum fluxes are
\begin{equation}
\sqrt{-g}T_t^z=-\omega\sqrt{-g}T_\varphi^z=\pm\frac{1}{x}A_{\varphi,x}B_T.
\end{equation}
On the axis, $\sqrt{-g}T_t^z\sim\mathcal{O}(1)$ or more regular, and asymptotically, $\sqrt{-g}T_t^z\sim\mathcal{O}(x^{-2})$ or more regular, for the above examples.

\subsubsection{pure AdS spacetime}

Keeping $l$ finite, we have for \eqref{ff_1st}
\begin{equation}
A_{\varphi,x}^2\mp\frac{l^2x}{2(l^2-x^2)}[B_T^2(x)]'=0,
\end{equation}
where \{$x\equiv\sin\theta/\sqrt{r^{-2}+l^{-2}},y\equiv r\cos\theta$\} such that similarly $\omega_\text{AdS}=\pm1/x$. Different from the flat case, here $0<x<l$ for $0<r<\infty$. Using a new coordinate $s\equiv\sqrt{1-x^2/l^2}$, the equation becomes simpler:
\begin{equation}\label{ff_1st_s}
A_{\varphi,s}^2\mp\frac{l^2}{2s}[B_T^2(s)]'=0,\qquad(1>s>0).
\end{equation}
A sample solution is given by (with the plus sign in \eqref{ff_1st_s})
\begin{equation}
l^2B_T^2(A_\varphi)=
\begin{cases}
cA_\varphi^n,\quad(n\neq2)\\
cA_\varphi^2
\end{cases},\qquad A_\varphi(s)=
\begin{cases}
\bigl[\frac{1}{2}cn(2-n)\ln s+C\bigr]^\frac{1}{2-n},\quad(n\neq2)\\
Cs^c
\end{cases}.
\end{equation}
These solutions are regular in the sense discussed for the flat case.

An analogy between the above solutions and a rotating monopole in the flat \cite{BZ} or AdS \cite{paper1} spacetime is that the former only depend on the coordinate $x$ on the 2-disc $(x,\varphi)$ and the latter only depend on $\theta$ on the 2-sphere $(\theta,\varphi)$, which manifests two different foliations of the spacetimes ($\text{2-disc}\times\mathbb{R}$ and $\text{2-sphere}\times\mathbb{R}$). 

For black hole cases ($m\neq0$), including NHEK, we have not found any solutions to the first-order equation. $\Omega^\pm$ is no longer monotonic in the radial direction and $x=1/\Omega^+$ is not suitable for a coordinate. However, the above solutions may be used as asymptotic configurations for black hole force-free magnetospheres.

\subsection{Special SAS solutions in NHEK spacetime}

The search for exact solutions in the near-horizon-extreme-Kerr (NHEK) background has been fruitful recently \cite{Strominger:NHEK_force-free,Lehner:NHEK_force-free}, making use of the underlying symmetries of the spacetime. Here we present some special SAS solutions in NHEK.
We work with the NHEK metric in global coordinates \cite{BardeenHorowitz:NHEK_global}
\begin{equation}\label{NHEK_global}
\operatorname{d}\!s^2=\frac{1+\cos^2\theta}{2}\Bigl[-(1+y^2)\operatorname{d}\!\tau^2+\frac{\operatorname{d}\!y^2}{1+y^2}+\operatorname{d}\!\theta^2\Bigr]+\frac{2\sin^2\theta}{1+\cos^2\theta}(\operatorname{d}\!\varphi^2+y\operatorname{d}\!\tau^2),
\end{equation}
and consider the following two cases.

\subsubsection{$A_\varphi=A_\varphi(y),B_T=B_T(y),\omega=\omega(y)$}\label{NHEK_1}

The force-free equation \eqref{dT_rth} becomes
\begin{equation}
f_1(y)\cos^4\theta+f_2(y)\cos^2\theta+[3f_1(y)-f_2(y)]=0,
\end{equation}
where
\begin{align}
f_1(y)&\equiv(y^2+1)A_\varphi'(y)\bigl[(y^2+1)A_\varphi''(y)+2yA_\varphi'(y)\bigr]+\frac{1}{2}[B_T^2(y)]'\\
\begin{split}
f_2(y)&\equiv2(y^2+1)A_\varphi'(y)\Bigl\{\bigl[2(\omega(y)+y)^2+y^2+1\bigr]A_\varphi''(y)+2\bigl[(\omega(y)+y)\omega'(y)+2\omega(y)+3y\bigr]A_\varphi'(y)\Bigr\}\\&\qquad+[B_T^2(y)]'.
\end{split}
\end{align}
Solving $f_1(y)=0=f_2(y)$ yields, after some simplifications,
\begin{align}
A_\varphi(y)&=\pm\int\frac{\sqrt{1-B_T^2(y)}}{y^2+1}\operatorname{d}\!y\\
\omega(y)&=-y\qquad\text{or}\qquad\omega(y)=\frac{y^2+1}{2\sqrt{1-B_T^2(y)}}\int\frac{y(y^2+1)[B_T^2(y)]'+4B_T^2(y)-4}{(y^2+1)^2\sqrt{1-B_T^2(y)}}\operatorname{d}\!y.
\end{align}
Then explicit forms of $B_T$ can be chosen manually. Some examples are listed below.

\begin{figure}[!tb]
\centering\includegraphics[scale=1.7]{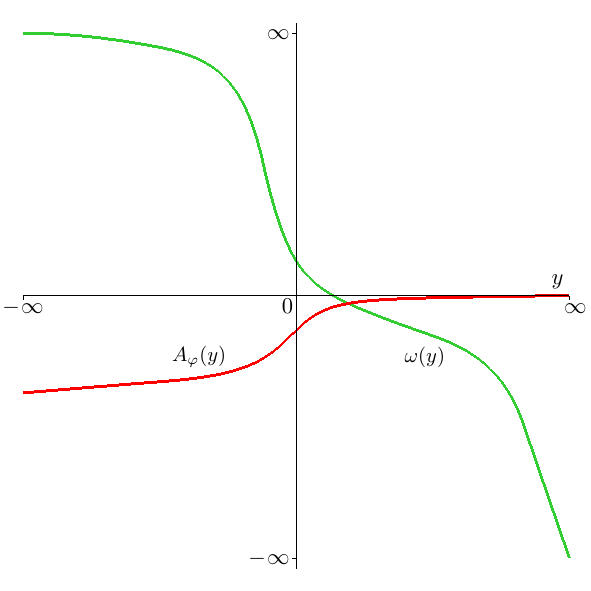}\includegraphics[scale=1.7]{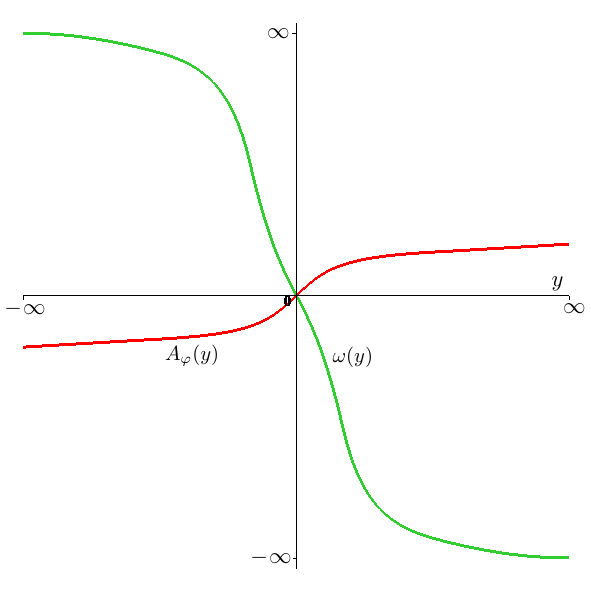}
\caption{Solutions for the first (left) and third (right) choices of $B_T$ in \cref{NHEK_1}. $A_\varphi(y)$: red curve, $\omega(y)$: green curve.}\label{NHEK_y}
\end{figure}

\begin{enumerate}
\item $B_T^2=2\arctan(y)/\pi$.
\begin{align}
A_\varphi(y)&=-\frac{\pi}{3}\Bigl[1-\frac{2\arctan(y)}{\pi}\Bigr]^\frac{3}{2}\\
\omega(y)&=-\frac{1}{6}(y^2+1)[3\sin(2\arctan(y))+4\arctan(y)-2\pi].
\end{align}

\item $B_T^2=1/(y^2+1)$.
\begin{align}
A_\varphi(y)&=\frac{1}{y}\\
\omega(y)&=-\frac{\text{sign}(y)}{\sqrt{(y^2+1)}}.
\end{align}

\item $B_T=B_T^c=\text{constant}$.
\begin{align}
A_\varphi(y)&=\sqrt{1-(B_T^c)^2}\arctan(y)\\
\omega(y)&=-y-(y^2+1)\arctan(y).
\end{align}
\end{enumerate}
Solutions in cases 1 \& 3 are plotted in \cref{NHEK_y}.

\subsubsection{$A_\varphi=A_\varphi(\theta),B_T=B_T(\theta),\omega=\omega(\theta)$}

The force-free equation \eqref{dT_rth} takes the form
\begin{equation}
(\dotso)y^2+(\dotso)y+\dotsc=0.
\end{equation}
Requiring vanishing of the coefficient of each power of $y$, we get
\begin{align}
A_\varphi'(u)&=\pm\frac{1+u^2}{1-u^2}\frac{1}{\sqrt{\omega(u)}}\\
[B_T^2(u)]'&=16\frac{\omega(u)^2+1}{\omega(u)}\frac{u(1+u^2)(3-u^2)}{(u^4+6u^2-3)(1-u^2)},\label{NHEK_u_BT_omega}
\end{align}
where $u\equiv\cos\theta$. One notices that there are two singular points in \eqref{NHEK_u_BT_omega} which cannot be easily removed by adjusting $\omega(u)$.

\section{Discussions}\label{disc}

A ubiquitous property of the solutions (for $\phi_1=0$ and $\phi_1=\phi_2=0$ configurations) found in the main text is the singular behavior on the horizon and/or rotation axis. It appears that to maintain the above assumptions on the field, the currents and/or energy densities diverge, indicating that the force-free condition may break down. (See \cite{MenonDermer_singular} for some similar observations.) The fact that one has to go to more special cases ($\Re\phi_1\text{ or }\Im\phi_1=0\longrightarrow\phi_1=0\longrightarrow\phi_1=\phi_2=0$) to find a (regular) force-free solution does raise questions as to which assumptions may be fruitful in searching for more general force-free solutions. 

In considering more general solutions, it is interesting to recall the existence of an argument \cite{ClassificationEM} that, given a suitable choice of null tetrad, non-null field configurations form a Ruse-Synge class with $\hat\phi_1\neq0$ \& $\hat\phi_{0,2}=0$ (where the hat denotes a special tetrad distinct from the one used in this paper). We leave a more detailed analysis of this possibility for future work. It would also be interesting to explore the use of duality rotations as a more general solution-generating technique.




\section*{Acknowledgment}

Special thanks to Dr.~Adam Ritz, for his valuable input to this paper via regular, in-depth discussions and review of the manuscript. Also to Dr.~Werner Israel for insights and comments.

\appendix

\section*{Appendix}

In this Appendix, we show that SAS solutions do not exist for Cases 0A, 0B, 1A, 1B \& 2A.

\subsection{Non-existence of SAS solutions for Case 0A: $\phi_0=\phi_2=0$}\label{append0A}

The force-free equations \eqref{FJ_eqns} impose $\Re\phi_1=0,J_m=0$ or $\Im\phi_1=0,J_l=J_n=0$. In either case, the vanishing of the corresponding current components requires $\partial_{r}(\phi_1/\rho^2)$ or $\partial_{\theta}(\phi_1/\rho^2)=0$ which is not possible.

\subsection{Non-existence of SAS solutions for Case 0B: $\phi_0=0$ or $\phi_2=0$}

With $\phi_2=0$, the degeneracy condition $\Im(\phi_1^2)=0$ implies two subcases:
\begin{enumerate}
\item $\Re\phi_1=0$, with the force-free equations \eqref{FJ_eqns} reducing to
\begin{equation}\label{ff0B1}
\Re(\bar\phi_0J_m)=0,\qquad 2i\Im\phi_1J_{\bar{m}}+\bar\phi_0J_n=0.
\end{equation}
The condition $J_n$ being real yields $\Im\phi_1=F_1(\theta)/\Sigma$. Then \eqref{ff0B1} impose $F_1=0$, so we get back to the null current configuration $\phi_2=\phi_1=0$.
\item $\Im\phi_1=0$, with the force-free equations \eqref{FJ_eqns} reducing to
\begin{equation}
J_n=0,\qquad\Re(\phi_1J_l-\bar\phi_0J_m)=0,
\end{equation}
while the first one is already impossible as argued in \cref{append0A}.
\end{enumerate}
Similar results hold for $\phi_0=0$.

\subsection{Non-existence of SAS solutions for Case 1A: $\Re\phi_1=0$}

The force-free equations reduce to (the first two being equivalent)
\begin{equation}\label{FJ_zero_Rephi1}
\Re(\bar\phi_0J_m)=0,\qquad 2i\Im\phi_1J_{\bar{m}}+\bar\phi_0J_n-\phi_2J_l=0,
\end{equation}
with \eqref{phi2'=c*phi1'} still holding (by the degeneracy condition). Then again have two subcases as in \eqref{2cases_SAS}:
\begin{enumerate}
\item  $\Im\phi'_0=0=\Im\phi_2'$

Solving the first force-free equation in \eqref{FJ_zero_Rephi1} and the conditions \eqref{dF=0_NP_J} yields
\begin{equation}
\phi_2'+\frac{1}{2}\phi_0'=\frac{\Xi}{\sqrt2}B_T=F_1(\theta),\qquad\phi_0'=\sqrt2a\Delta_\theta\sin^2\theta\frac{F_2(\theta)}{\Sigma}+F_3(\theta),\qquad\Im\phi_1=\frac{F_2(\theta)}{\Sigma},
\end{equation}
for arbitrary $F_{1,2,3}(\theta)$. However, using these to evaluate the second force-free equation necessarily leads to $F_2(\theta)=0$, bringing us back to Case 0C in \cref{case1}.
\item $c=\Delta_r/(2\Sigma)$

Solving the imaginary part of the first force-free equation in \eqref{FJ_zero_Rephi1} leads to $\Im\phi_1=0$ or  $\Im\phi_0'=0$ (i.e., the previous subcase which again implies $\Im\phi_1=0$), and thus back to Case 0C in \cref{case1}.
\end{enumerate}

\subsection{Non-existence of SAS solutions for Case 1B: $\Im\phi_1=0$}

The force-free equations reduce to (the first two being equivalent)
\begin{equation}\label{FJ_zero_Imphi1}
\Re(\phi_1J_n-\phi_2J_m)=0 
,\qquad\frac{J_n}{J_l}=\frac{\phi_2}{\bar\phi_0},
\end{equation}
and again we have \eqref{phi2'=c*phi1'} and two subcases given by \eqref{2cases_SAS}:
\begin{enumerate}
\item $\Im\phi'_0=0=\Im\phi_2'$

Evaluating the reality condition of $J_n$ leads to $\Re\phi_1=0$ and thus back to Case 0C in \cref{case1}.
\item $c=\Delta_r/(2\Sigma)$

Solving the second force-free equation in \eqref{FJ_zero_Imphi1} and the conditions \eqref{dF=0_NP_J} yields
\begin{equation}
\Re\phi_1=\frac{F_1(r)}{\Sigma},\qquad e^R{(r,\theta)}=\frac{\sec[\Theta(r,\theta)]}{F_2(r)},\qquad\tan[\Theta(r,\theta)]=-\frac{\sqrt2\Delta_r}{a\Sigma}F_1(r)F_2(r)+F_3(r),
\end{equation}
and then evaluating the remaining first force-free condition imposes $F_1(r)=0$ so again we get back to Case 0C in \cref{case1}.
\end{enumerate}

\subsection{Non-existence of SAS solutions for Case 2A: $\Re(\phi_1^2)=-\Re(\Phi_2^2)/(8\Delta_r)$}

Without loss of generality, we set $R_2=0$ using the notations in \cref{othercases}. One can derive from the conditions \eqref{dF=0_NP_J} explicit forms of $\Theta(r,\theta),R_1(r,\theta)$ which we omit but note that $\tan\Theta\neq1$ so the special subcase in Case 2A is excluded. To simplify matters, we now look at the $a=0$ limit, and notice that the above found $R_1$ and that from solving the force-free equations do not agree, their ratio being
\begin{equation}
\sqrt{\csc^2\theta(1+l^2r^{-2}-2ml^2r^{-3})-(1+4ml^2r^{-3}-9m^2l^2r^{-4})}\,\text{function}(1+l^2r^{-2}-2ml^2r^{-3})\neq1,
\end{equation}
which implies there are no solutions for nonzero $a$ (though a solution exists in the $m=a=0$ limit).


\bibliography{paper2-bib}

\end{document}